\documentclass[sigconf,natbib=true,anonymous=false]{acmart}
\usepackage{makecell}
\AtBeginDocument{%
  \providecommand\BibTeX{{%
    \normalfont B\kern-0.5em{\scshape i\kern-0.25em b}\kern-0.8em\TeX}}}

\acmConference[Workshop on Multimodal Representation and Retrieval at SIGIR 2024]{July 14--18,
  2024}{Washington, DC}
\begin{document}

\title{Async Learned User Embeddings for Ads Delivery Optimization}

\author{Mingwei Tang*, Meng Liu, Hong Li, Junjie Yang, Chenglin Wei, Boyang Li, Dai Li, Rengan Xu, Yifan Xu, Zehua Zhang, Xiangyu Wang, Linfeng Liu, Yuelei Xie, Chengye Liu, Labib Fawaz, Li Li, Hongnan Wang, Bill Zhu, Sri Reddy}
\email{Corresponding_author: *mingwt@meta.com}
\affiliation{%
   \institution{Meta Platforms, Inc}
   \country{United States}
}

\renewcommand{\shortauthors}{Tang and Liu, et al.}

\begin{abstract}
In recommendation systems, high-quality user embeddings can capture subtle preferences, enable precise similarity calculations, and adapt to changing preferences over time to maintain relevance. The effectiveness of recommendation systems depends on the quality of user embedding. We propose to asynchronously learn high fidelity user embeddings for billions of users each day from sequence based multimodal user activities through a Transformer-like large scale feature learning module. The async learned user representations embeddings (ALURE) are further converted to user similarity graphs through graph learning and then combined with user realtime activities to retrieval highly related ads candidates for the ads delivery system. Our method shows significant gains in both offline and online experiments. 
\end{abstract}

\ccsdesc[500]{Information systems~Top-k retrieval in databases}
\ccsdesc[500]{Information systems~Similarity measures}
\ccsdesc[500]{Information systems~Novelty in information retrieval}
\ccsdesc[300]{Information systems~Content match advertising}
\ccsdesc[100]{Information systems~Social networks}
\ccsdesc[500]{Information systems~Personalization}


\keywords{Transformer, multimodal, user representation, retrieval, graph}


\received{25 April 2024}
\received[accepted]{23 May 2024}

\maketitle

\section{Introduction}
High fidelity user representation lies at the core of recommendation retrieval and ranking \cite{pal2020pinnersage, baltescu2022itemsage, pi2019practice, grbovic2018real, zhang2023scaling, tencent, yuan}. By accurately capturing user preferences, behaviors, and interactions, these representations empower recommendation systems to deliver personalized, relevant, and timely recommendations that enhance user satisfaction and engagement. User representations condense the complex user information into low-dimensional vectors.The quality of user embedding directly impacts the effectiveness and accuracy of recommendation systems. High-fidelity user embeddings can capture nuanced user preferences and contribute to more precise similarity computations, e.g. identify similar users and recommend ads that are engaged by the similar users. Recommendation systems also need to adapt to user preference changes over time to maintain relevance and effectiveness \cite{zhang2023scaling, pi2019practice}. High-quality user embeddings capture temporal dynamics and evolving user preferences, enabling recommendation models to continuously update and refine user representations based on the most recent interactions and feedback. Recent advances in Transformer based structures \cite{xia2023transact} have demonstrated its effectiveness in capturing both the content and temporal information from sequential inputs and graph based learning algorithms are also proven to be useful in improving recommendation systems \cite{wu2022graph, yang2018graph, wang2021graph, darban2022ghrs}, which motivates us to combine a custom Transformer-like structure with graph learning to improve user representations and ads delivery system.

\begin{figure}[h]
  \centering
  \includegraphics[width=0.8\linewidth]{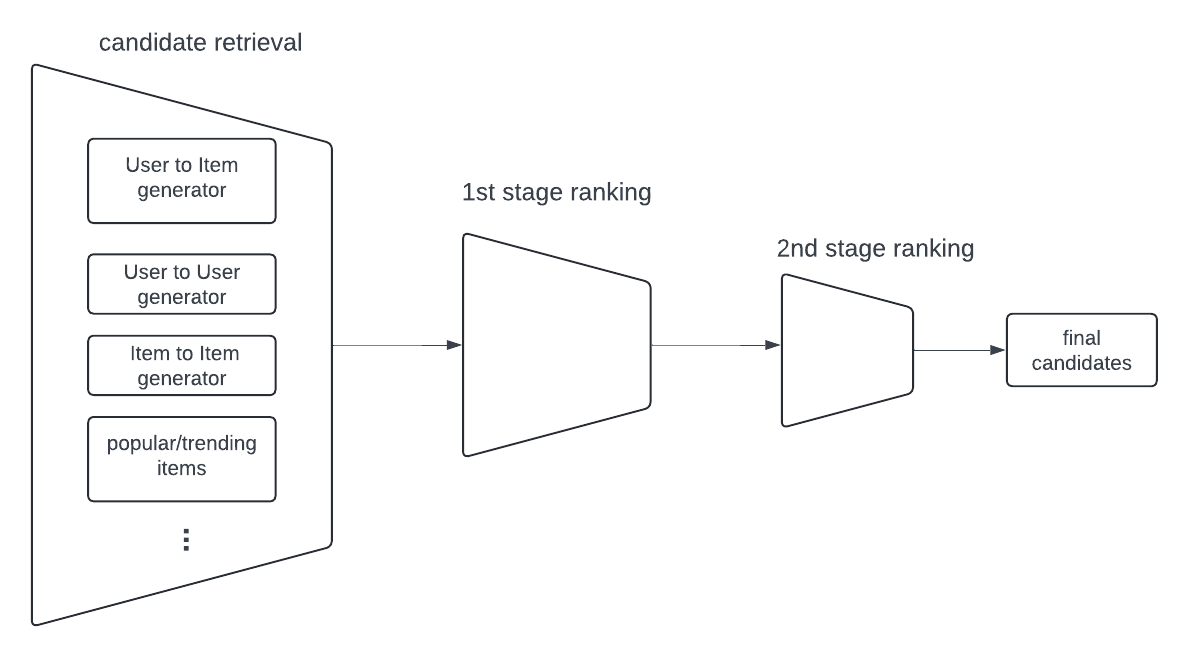}
  \caption{Ads delivery funnel with multi-stage ranking}
  
  \label{fig:multistage}
\end{figure}

\begin{figure}[h]
  \centering
  \includegraphics[width=0.8\linewidth]{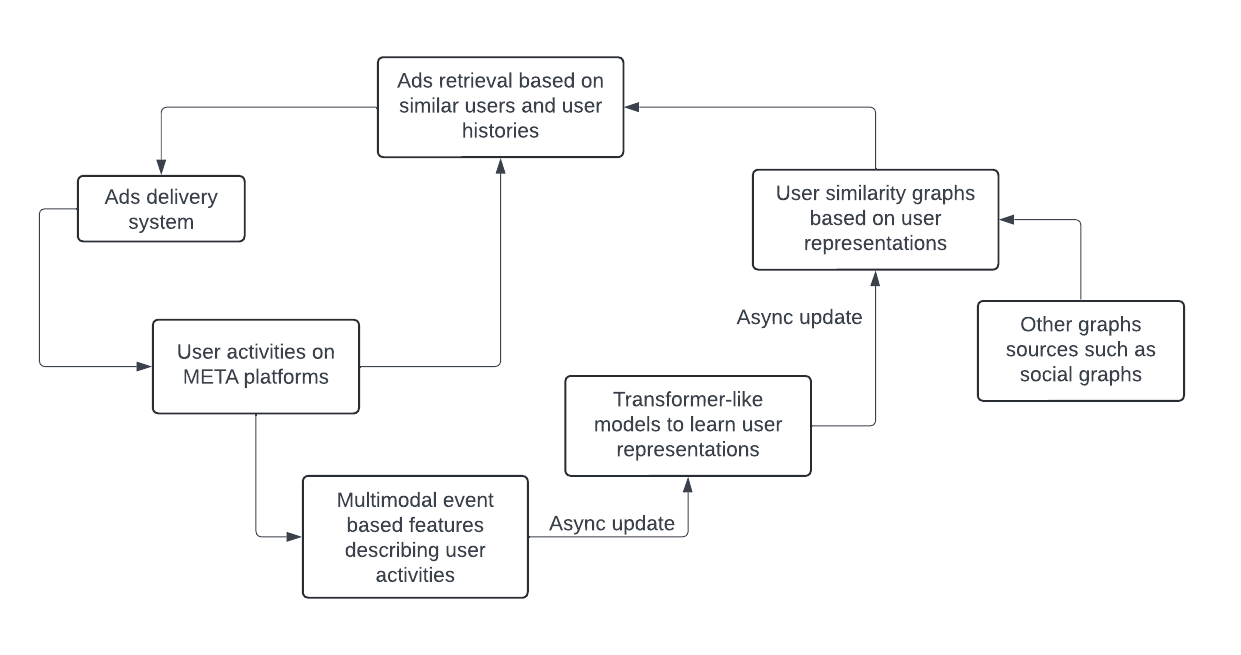}
  \caption{The feedback loop for User Representation Learning}
  
  \label{fig:overview}
\end{figure}

In this paper, we introduce how to generate high fidelity compact user representations and leverage it to improve retrieval performance. In industry, ads are delivered through a multi-stage ranking system \cite{wang2023towards, covington2016deep} as shown in Figure \ref{fig:multistage}. 
Our user representations can be used in all stages, but we will mainly focus on improving the candidate retrieval stage with a user-to-user (u2u) generator retrieving ads based on similar users. At a high level, we propose to improve the retrieval stage with the following components (see Figure \ref{fig:overview}):
\begin{itemize}
    \item Multimodal event based features describing various user activities, such as user comments, user clicked ads or user viewed pictures and short form videos.
    \item A custom designed Transformer-like model to fuse multimodal sequence based inputs into a few high fidelity user embeddings. These embeddings are async updated to support large scale models up to over 10 GFLOPS.
    \item User similarity graphs based on user representations. These graphs are also async updated to support billions of users.
    \item During ads delivery, ads candidates are retrieved based on a combination of user similarity graph and other graphs sources such as social graphs, as well as user historical interactions with ads.
\end{itemize}

\section{Async Learned User Representation Embedding}
In this section, we will explain how we leverage multimodal user history inputs to compute user embeddings in async fashion.
\subsection{Multimodal and sequence based user history inputs}
The proposed system utilizes multimodal user history events from various sources. Representative event based features include user clicked ads ids or metadata information such as categories, encoded user generated texts such as comments and Residual-Quantized Variational Autoencoder (RQ-VAE) \cite{lee2022autoregressive} encoded user viewed images or short form videos.

\begin{figure}[h]
  \centering
  \includegraphics[width=0.8\linewidth]{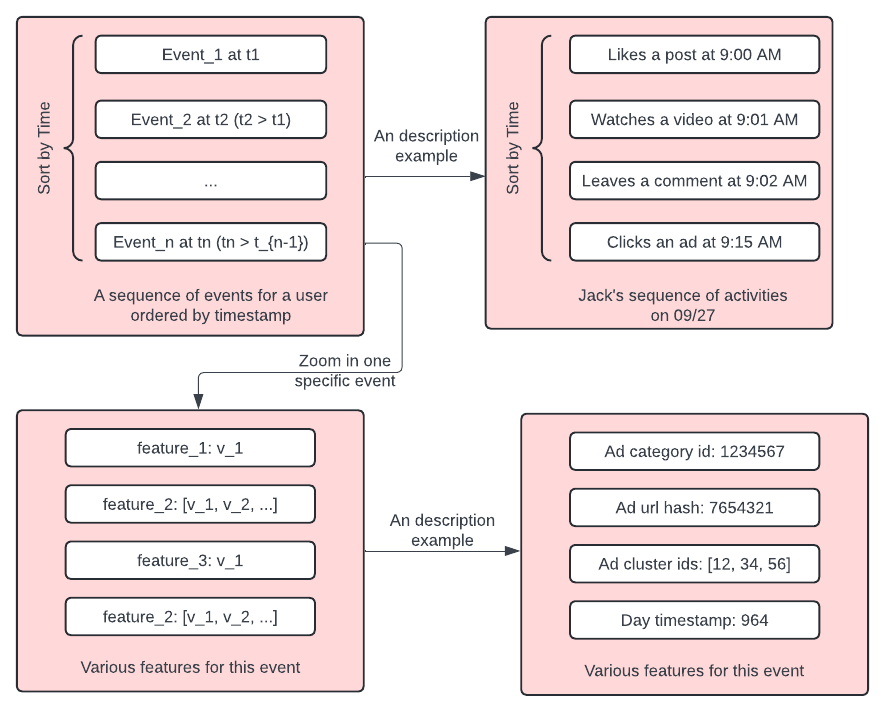}
  \caption{An example of user sequence feature}
  \label{fig:EBF}
\end{figure}

These signals are encoded as sequential data together with the corresponding timestamp for each event. An example of sequence based feature is shown in Figure \ref{fig:EBF}.

\subsection{Customized Transformer-like user representation module}
\begin{figure}[h]
  \centering
  \includegraphics[width=\linewidth]{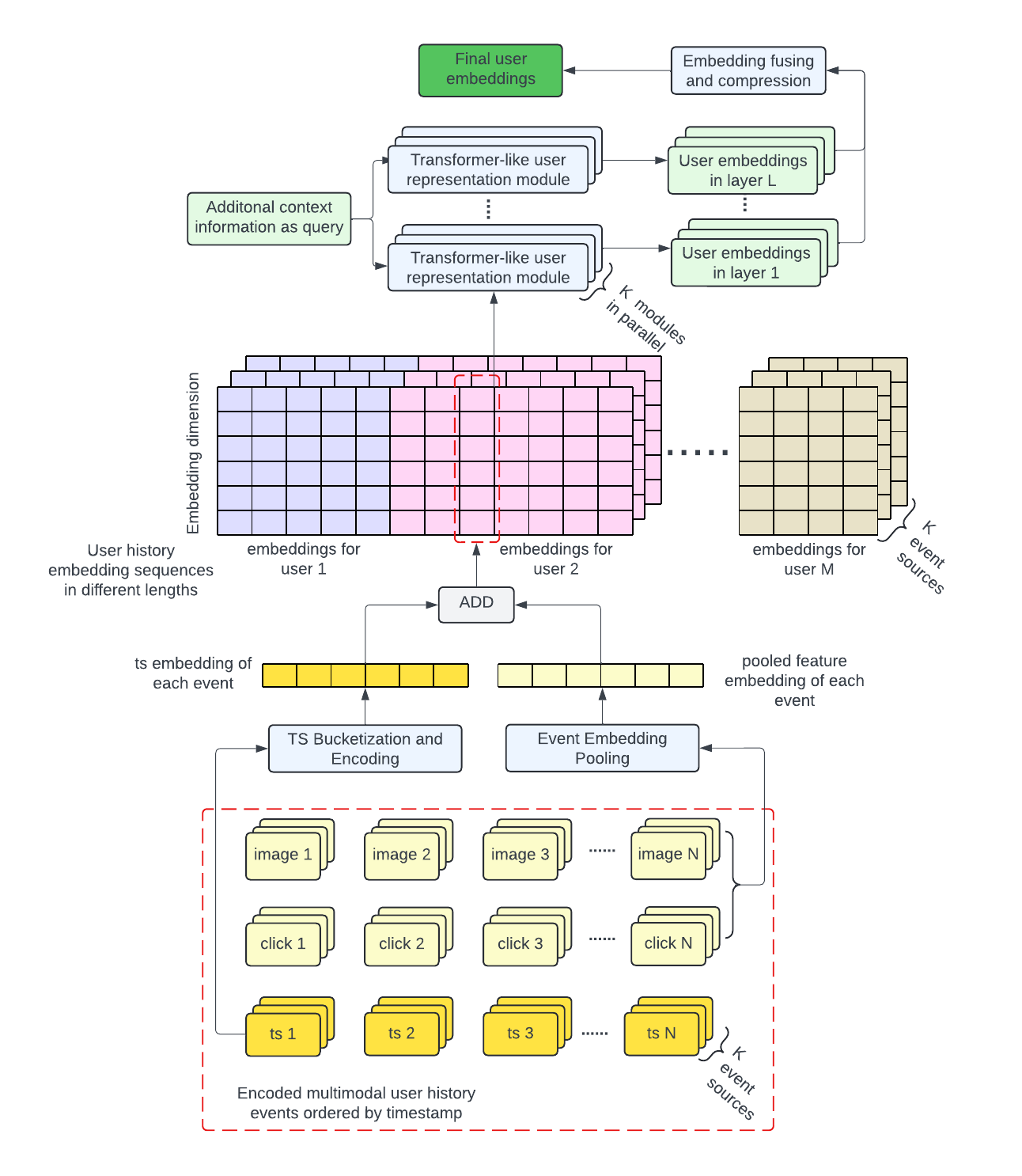}
  \caption{Illustration of user representation embedding learning architecture}
  
  \label{fig:user_embeddings}
\end{figure}

To better process the sequential data and learn the higher order interactions among different event embeddings, we design a multi-layer Transformer-like user representation module. To better capture multimodal information from $K$ different data sources, at each layer, we use $K$ independent modules in parallel, one module for each source. User embeddings can be extracted from any intermediate layers of our multi-layer user representation module. (See Figure \ref{fig:user_embeddings})

\begin{figure}[h]
  \centering
  \includegraphics[width=\linewidth]{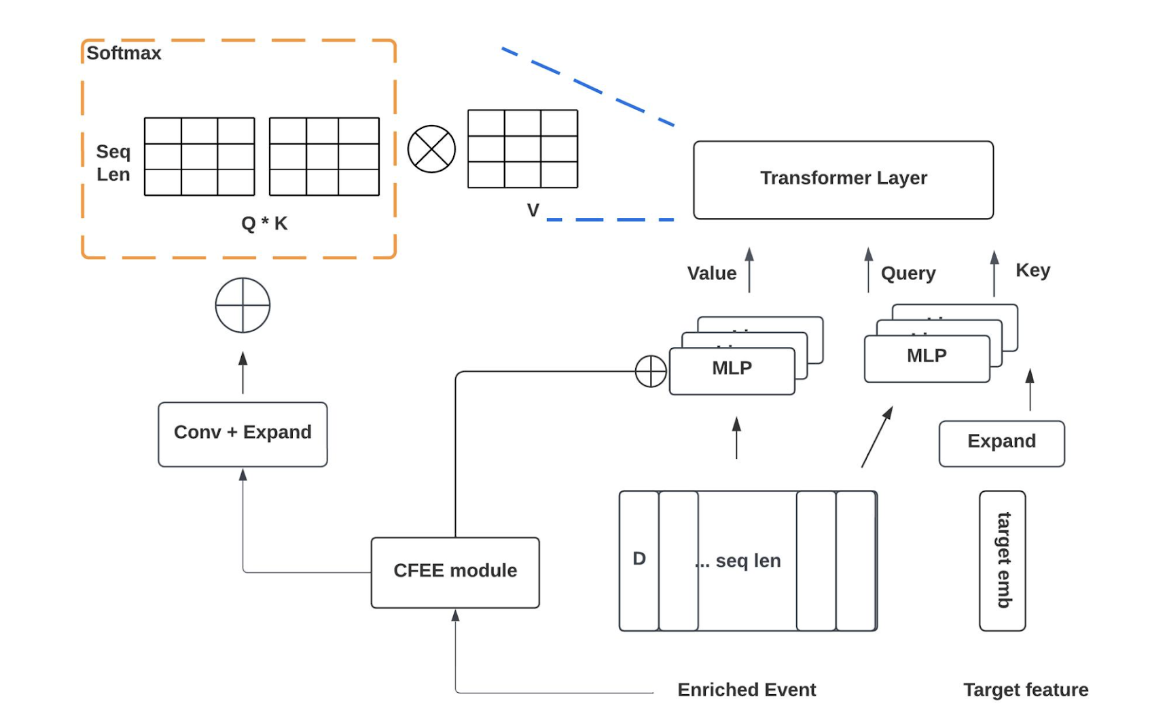}
  \caption{One layer of our custom Transformer-like architecture}
  
  \label{fig:TU_details}
\end{figure}

Figure \ref{fig:TU_details} shows the detailed design of each Transformer-like user representation module. Compared to the traditional Transformer implementation, our modules have several key differences
\begin{itemize}
\item Our module can handle user history sequences in different lengths without any padding. This can help scale our system to include user history information from months or even years ago.
\item Besides the standard position encoding used in standard Transformer, our module uses a complex feature enrichment encoder (CFEE), which takes the raw timestamp sequence as inputs and encodes it through various encoding methods. Our current CFEE module includes:
\begin{itemize}
\item Absolute position encoder \cite{vaswani2017attention}. We apply absolute position encoder to information such as user engagement frequencies and user engagement order.
\item Temporal decay encoder \cite{pancha2022pinnerformer}. To model the importance of the past user ads engagement events decreasing over time, we apply the temporal decay encoder using a logarithmic transformation of time.
\item Cyclic pattern encoder \cite{su2021roformer}: user ads engagement shows a cyclical pattern. For instance, the level of engagement varies throughout the day, with different engagement densities at different times. Additionally, the engagement pattern differs between weekdays and weekends.. We adopt such idea and use a similar encoder to capture the cyclic patterns.
\item Relative position encoder \cite{shaw2018self}: Calculating relative distance poses a challenge in neural networks because position indices become invisible within the network after vector embedding. Discrete position indices must be embedded as vectors in order to be incorporated into the gradient during backpropagation.
\end{itemize}
\item Unlike standard Transformer, where the encoded information is directly added to the content embeddings, we also compute pairwise relative timestamp encoding through CFEE module and add it to the attention weights matrix as attention bias.
\item  Besides the standard self attention, our module can also take additional context information such as ads embeddings as queries to further learn the interactions between users and ads.
\end{itemize}

\subsection{Effective embedding compression}
By default, the Transformer-like user representation module will produce hundreds of user embeddings, hence an effective embedding compression arch is designed to compress them into just a few to achieve storage saving and computation saving in the later stage of building user graphs, as well as fusing embeddings from different layers and sources. Our compression arch can either (1) use ResNet \cite{he2016deep} style skip connection to combine compressed information from dot product and linear compressed user embeddings (2) use another Transformer-like interaction module.

\subsection{Async updates and user embedding logging}
To provide better scalability, the user embeddings are not computed in realtime, but precomputed and updated in async fashion. During evaluation, we will not evaluate the entire model. Instead, we only evaluate the parts that are responsible for user embedding generation (see red box in Figure \ref{fig:AFL_process}). This can either increase our evaluation speed by ~50\% or reduce the number of GPUs needed by 75\%. Depending on the complexity of the model, the refresh latency can be tuned from minutes to days based on the application, infra capacity and other considerations.

\section{User Similarity Construction}

\begin{figure}[h]
  \centering
  \includegraphics[width=0.8\linewidth]{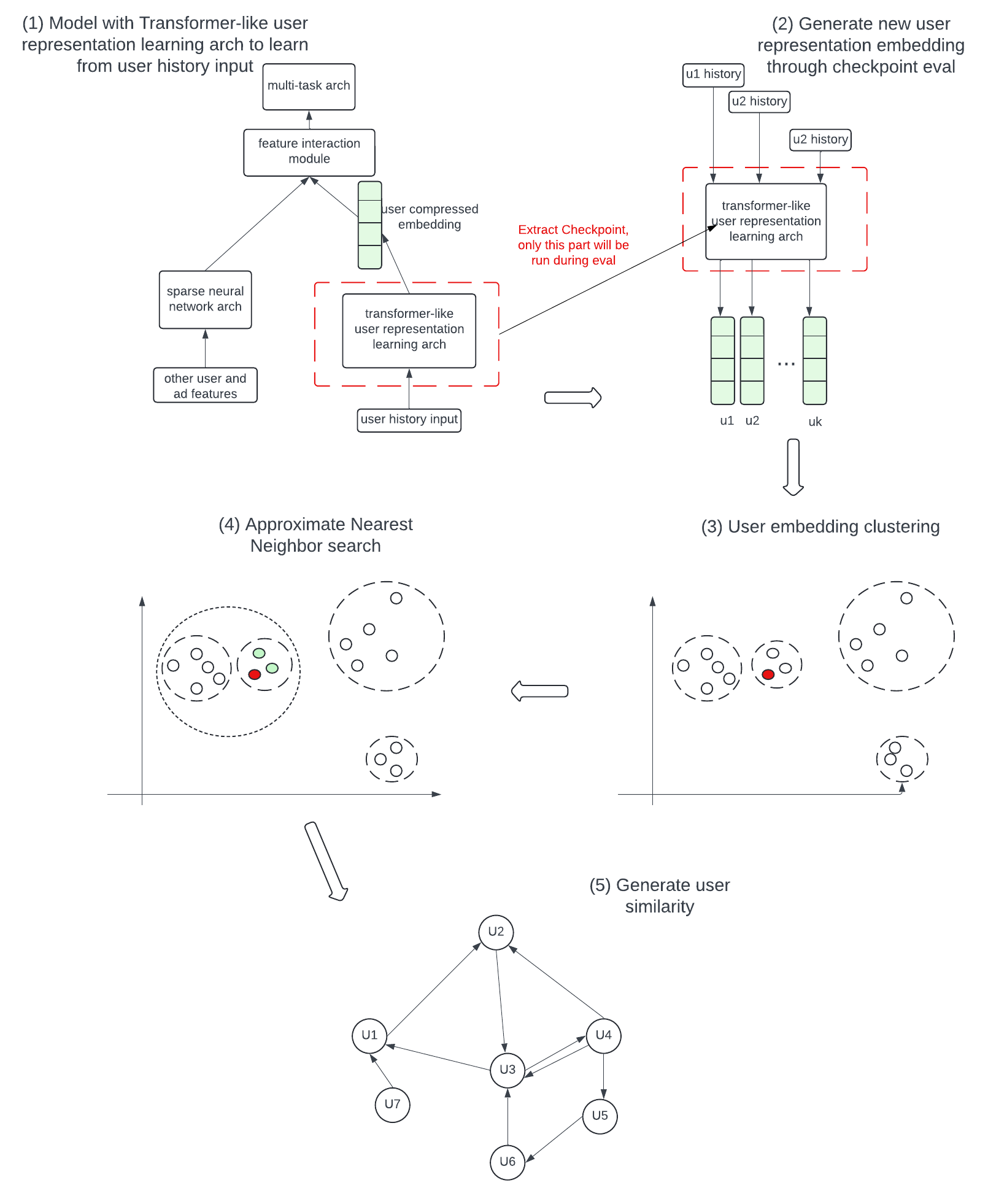}
  \caption{User similarity construction through user representation embedding learning}
  
  \label{fig:AFL_process}
\end{figure}

This section explains how to construct user similarity graph from async learned user representation embeddings (ALURE). User similarity is characterized by cosine distance in the user embedding space. Figure \ref{fig:AFL_process} provides an illustration of the end-to-end process.
\begin{itemize}
    \item A model with a Transformer-like feature arch performs recurring training using the training data with user history input. New model checkpoints are generated on a daily basis. 
    \item The Transformer-like feature arch is extracted from the model checkpoint to perform evaluation using new user history input data, which eventually generates user representation embeddings. 
    \item We reduce the similar user search space using clustering:
    \begin{itemize}
        \item Users within the same country will be clustered through K-means method with $k_1$ clusters based on the cosine distance of in user representation embedding space.
        \item The search space is reduced to a small subset of $k_1’$ ($k_1’ << k_1$) nearest clusters based on the cosine distance to the cluster centroid. 
    \end{itemize}
    \item We search $k_2$ nearest neighbors from the $k_1'$ nearest clusters. The FAISS KNN \cite{douze2024faiss} library is applied to accelerate to nearest neighbor search.
        \item A direct graph is constructed by forming edges from users to their corresponding nearest neighbors. The graph construction pipeline will be executed daily to reflect the latest user behaviors. 
\end{itemize}

\section{Improve Ads Delivery with User Graphs}
\begin{figure}[h]
  \centering
  \includegraphics[width=0.8\linewidth]{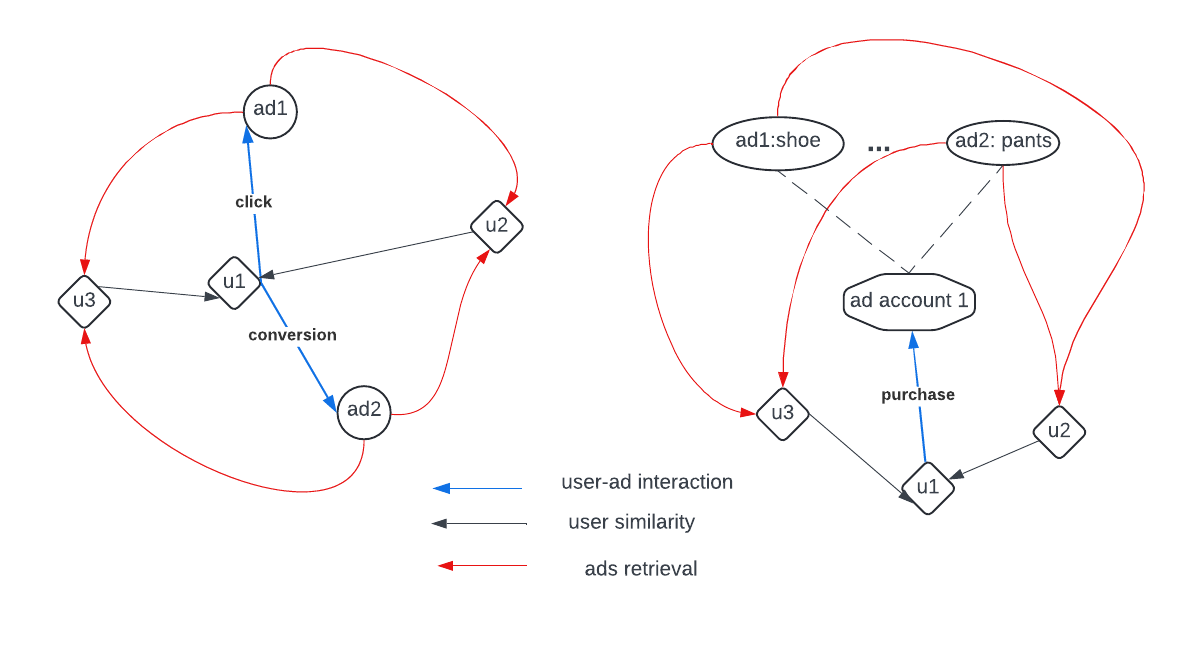}
  \caption{Retrieval Related Ads to similar users}
  
  \label{fig:delivery_optimization}
\end{figure}

The similarity graph from ALURE is used as a u2u generator at the candidate retrieval stage in figure \ref{fig:multistage}. In this process, related ads from one user’s engagement history will be retrieved as the ads candidate for his/her similar users at the retrieval stage. Related ads can be
\begin{itemize}
    \item the exact ads that the user has click/conversion behaviors 
    \item a list of relevant ads under the ad account where the user has engaged with that ad account
\end{itemize}
Figure \ref{fig:delivery_optimization} on the left shows the first scenario, where the ad1 and ad2 that have been clicked by u1 is retrieved as ads candidates for similar users u2 and u3. Figure \ref{fig:delivery_optimization} on the right depicts the second scenario in which u1 has a conversion event related to the ad or product under an account, for example, purchasing a T-shirt. Then other ads under the same ad account like shoe or pants can be retrieved to similar users u2 and u3.  

\section{Experiments}

\begin{table}[h]
  \caption{Model improvement from using asynchronous learned user representation embedding (ALURE) as feature}
  \label{tab:freq}
  \begin{tabular}{ccl}
    \toprule
    Model type&Prediction task&NE gain\\
    \midrule
    Mobile feed & CTR & 0.12\%\\
    Mobile feed & CVR & 0.37\%\\
    Instagram & CTR & 0.1\%\\
  \bottomrule
\end{tabular}
\end{table}

\subsection{Results in offline experiments}
We first verify the quality of generated user embeddings by directly using them as features in production ads ranking models. In our experiments, we find that these embeddings can provide $0.1 \sim 0.3\%$ Normalized Cross Entropy (NE) statistically significant gain on important tasks of production ads ranking models such as CTR or CVR, proving the effectiveness of these embeddings.

\subsection{Results in online experiments}
\begin{figure}[h]
  \centering
  \includegraphics[width=0.8\linewidth]{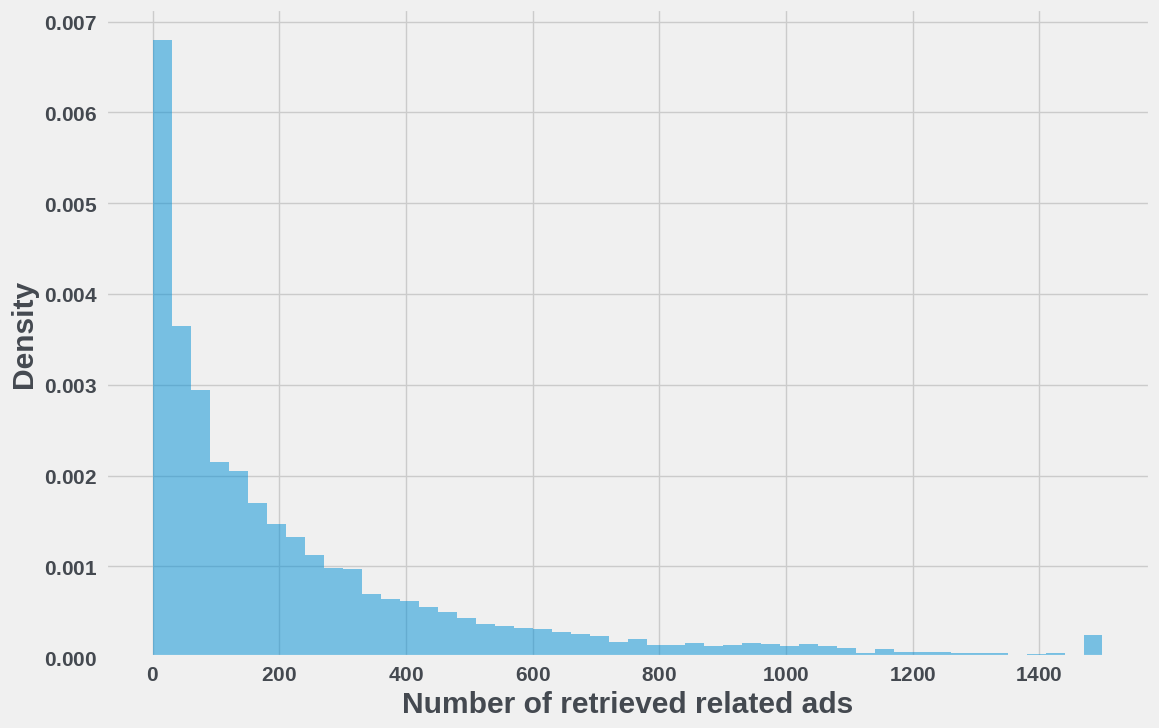}
  \caption{Distribution of user level number of retrieved related ads}
  
  \label{fig:ads_dist}
\end{figure}

We set up online A/B testing experiments to measure the performance of ALURE in delivery optimization. The experiment consists of one control version reflecting the current system and two test versions with additional u2u generators at the retrieval stage. In test version 1, we use (1) user similarity based on user following relationship (BFF) and (2) user similarity learned from Personalized PageRank (PPR). Version 2 is built on the top of version 1 with additional ads retrieved based on user graph using ALURE. The user graph is constructed by searching $k_2=15$ similar users from $k_1’=50$ nearest clusters out of the $k_1 = 400$ clusters from K-means. 
The total number of retrieved ads for each user from ALURE is capped to 1500. Figure \ref{fig:ads_dist} demonstrates the distribution of the retrieved related ads for each user.

\begin{table}[h]
 \centering
  \caption{Online performance from using ALURE for delivery optimizaiton}
  \label{tab:online}
  \begin{tabular}{ccc}
    \toprule
    \makecell{Experiment \\ version}&\makecell{User similarity \\ graph}&\makecell{Online metric \\ change}\\
    \midrule
    control&  & 0\% \\
    Version 1&BFF + PPR & -0.05\% \\
    Version 2&\makecell{BFF + PPR + \\ ALURE} & $\boldsymbol{0.28\%}$ \\
  \bottomrule
\end{tabular}
\end{table}

The performance is measured with on an online metric calculated from the total value generated from both advertisers and people based on the ads shown and engaged with. We report the relative online metric change using 
\begin{equation}
    (metric_{test} - metric_{control}) / metric_{control} \times 100\%.
\end{equation}

Version 1 with BFF and PPR graph shows -0.05\% online metric change compared with the control group, which is within the $\pm 0.15\%$ statistical significant threshold and should be considered as neutral. Version 2 with additional ads from ALURE demonstrates 0.28\% statistically significant improvement on ads online metric, indicating the strong performance of ALURE in ads delivery optimization.

\section{Acknowledgments}
This work would not be possible without collaboration from hundreds of people. Especially, we want to thank (in alphabetical order) Baichuan Yuan, Ben Wang, Dianel Li, Doris Wang, Esam Abdel Rhman, Hong Yan, Hung Duong, Jiaqi Zhai, Joy Mu, Lang Yu, Leo Ding, Pan Chen, Peng Sun, Pushkar Tripathi, Qiao Yang, Qian Ge, Rocky Liu, Santanu Kolay, Sandeep Pandey, Xiaoyu Chen, Xing Liu, Xuanting Cai.
\section{Summary}

\bibliographystyle{ACM-Reference-Format}
\bibliography{sample-base}










\end{document}